\documentclass{elsart}

\usepackage{amssymb}
\usepackage{color}
\usepackage{graphicx}

\begin{document}

\begin{frontmatter}

%

\title{Domain imaging, MOKE and magnetoresistance studies of CoFeB films for MRAM applications}


\author[Porto]{J.M. Teixeira}
\author[Porto]{R.F.A. Silva}
\author[Porto]{J. Ventura}
\author[Porto]{A.M. Pereira}
\author[Porto]{F. Carpinteiro}
\author[Porto]{J.P. Ara\'ujo}
\author[Porto]{J.B. Sousa\corauthref{cor}}
\corauth[cor]{Corresponding author. Tel.: +351226082656; fax:
+351226082679} \ead{jbsousa@fc.up.pt}
\author[Lisboa]{S. Cardoso}
\author[Lisboa]{R. Ferreira}
\author[Lisboa]{P.P. Freitas}

\address[Porto]{DFFCUP and IFIMUP, Rua do Campo Alegre, 687, 4169-007 Porto, Portugal}
\address[Lisboa]{IST and INESC-MN, Rua Alves Redol, 9-1, 1000-029 Lisbon, Portugal}

\begin{abstract}
We present a detailed study on domain imaging, Kerr effect
magnetometry (MOKE) and magnetoresistance (MR), for a series of 20
nm Co$_{73.8}$Fe$_{16.2}$B$_{10}$ thin films, both as-deposited
(amorphous) and annealed (crystalline). By considering the two
different (orthogonal) in-plane magnetization components, obtained
by MOKE measurements, we were able to study the uniaxial anisotropy
induced during CoFeB-deposition and to discriminate the
magnetization processes under a magnetic field parallel and
perpendicular to such axis. MOKE magnetic imaging enabled us to
observe the dominant magnetization processes, namely domain wall
motion and moment rotation. These processes were correlated with the
behavior of the magnetoresistance, which depends both on short-range
spin disorder electron scattering and on the angle between the
electrical current and the spontaneous magnetization
($\emph{\textbf{M}}_{S}$). A simple numerical treatment based on
Stoner-Wolfarth model enables us to satisfactorily predict the
magnetization behaviour observed in these films. A comparison
between the results in Co$_{73.8}$Fe$_{16.2}$B$_{10}$ films and the
previous ones obtained in annealed Co$_{80}$Fe$_{20}$ films, show
that the introduction of boron in CoFe reduces significatively the
coercive and saturation fields along the easy axis (e.g. $H_{c}$
from $\sim$ 2 down to $\sim$ 0.5 kAm$^{-1}$). Also, the
magnetization along the hard axis saturates at lower fields. We
conclude that amorphous and nanocrystalline CoFeB films show low
coercive fields and abrupt switching, as well as absence of short
range spin disorder effects after switching when compared with
Co$_{80}$Fe$_{20}$.
\end{abstract}

\end{frontmatter}

%
\section{Introduction}
\label{introduction} Tunnel junctions (TJ) consisting of two
ferromagnetic (FM) layers separated by an insulator \cite{Modera}
are strong candidates for leading technological applications such
as sensor elements in read heads \cite{readheads} and non-volatile
magnetic random-access memories (MRAMs) \cite{MRAMs}. The
magnetization of one of the FM layers (pinned layer) is fixed by
an underlying antiferromagnetic (AFM) layer, whereas the
magnetization of the other FM layer (free layer) reverses almost
freely when a small magnetic field is applied. Due to spin
dependent electron tunneling one can thus have two distinct
resistance (R) states, associated with the magnetizations of the
pinned and free layers parallel (low R) or antiparallel (high R).
To improve device performance, one continuously aims to achieve
higher tunnel magnetoresistance (TMR), better thermal stability
and low ferromagnetic coupling ($H_f$) between pinned and free
layers. The use of amorphous CoFeB films in the free and pinned
layers of optimized tunnel junctions enabled us to obtain a TMR
coefficient as high as 70\% \cite{CoFeB70}, good transport
properties upon annealing up to 673 K \cite{thermalstability} and
coercive and coupling fields as low as $\sim$ 160 Am$^{-1}$
\cite{CoFeB}.

Here we present a study on the magnetoresistance (MR), Kerr Effect
vectorial magnetometry and domain imaging of a series of 20 nm
(Co$_{73.8}$Fe$_{16.2}$)B$_{10}$ films, both as-deposited and
annealed. A CCD camera with $\sim$ 10 $\mu$m resolution enabled
direct domain visualization. X-ray diffraction showed that the
annealed films were crystalline, with a strong (110) texture,
while the as-deposited CoFeB films were amorphous \cite{CoFeB}.

By considering the two different in-plane magnetization (M)
components (given by vectorial MOKE magnetometry), we study the
M-processes under longitudinal (easy axis) and transverse (hard
axis) fields (H). The results are compared with simultaneous MR
measurements and magnetic domain visualization. Under transverse
fields magnetic rotation processes dominate and lead to good
correlation between the behavior of M, MR and domain changes under
H. For longitudinal fields the M-processes are mainly due to
domain wall displacements ($180^{\circ}$ magnetization reversals)
and no MR dependence is observed in this case, both for the
annealed and amorphous CoFeB films. Thus, the uniaxial anisotropy
is responsible for the different magnetization reversal processes
observed when the \textbf{\emph{H}} is applied along the easy or
hard axes.


\section{Experimental Details}
\label{Experimentaldetails} We studied a series of as-deposited
and annealed (10 min. at 553 K) (Co$_{73.8}$Fe$_{16.2}$)B$_{10}$
thin rectangular films (4 mm $\times$ 4 mm $\times$ 20 nm for the
as-deposited sample and 3 mm $\times$ 14 mm $\times$ 20 nm for the
annealed sample) grown on glass substrates by ion-beam deposition
\cite{CoFeB}. A magnetic field of $240 \times 10^{3}$ Am$^{-1}$
was applied along the longitudinal direction during deposition,
inducing an easy axis direction in all the studied films. The
magnetic properties were investigated at room temperature by
Magneto-Optical Kerr effect (MOKE), domain imaging and MR
measurements \cite{Teixeira}. MOKE hysteretic cycles were obtained
using a vectorial MOKE magnetometry unit simultaneously measuring
both in-plane magnetization components, allowing us to obtain the
technical magnetization vector \textbf{\emph{M(H)}} in the two
common in-plane geometries: the \emph{transverse} geometry, with
\textbf{\emph{H}} in the film plane and perpendicular to the
laser-beam incident plane; and the \emph{longitudinal} geometry,
in which the in-plane field is parallel to the incident plane. A
greyscale CCD camera with $\sim10 \mu$m of resolution is used to
acquire the magnetic domain images. Each image is saved in a
bitmap format with 8 bit of information and is then differentiated
with respect to the magnetically saturated image to enhance image
contrast. In both systems the sample is located in the center of a
pair of Helmholtz coils.

The four probe technique was used for the MR measurements, with
the electric current along the long axis of our rectangular films
and the in-plane applied magnetic field parallel or at right
angles to the electrical current. In ferromagnetic 3d-transition
metals the electric resistivity depends on the angle $\theta$
between the electrical current and the spontaneous magnetization
$\emph{\textbf{M}}_{S}$, through the so called anisotropic
magnetoresistive effect (Smit mechanism; see \cite{AMR}):
\begin{equation}\label{Definição-de-R}
    \rho(\textbf{\emph{H}})=\rho_\perp+(\rho_\parallel-\rho_\perp)\cos^2\theta,
\end{equation}
where $\rho_{\perp}(\rho_{//})$ is the resistivity when
\emph{\textbf{M}} is saturated perpendicular (parallel) to the
electrical current. A magnetoresistive coefficient (at field
\emph{\textbf{H}}) is defined as:
\begin{equation}\label{Definição-de-AMR}
    \frac{\Delta\rho}{\rho}=\frac{\rho(\emph{\textbf{H}})-\rho(0)}{\rho(0)}.
\end{equation}
For a film with $\emph{\textbf{M}}_{S}$ always in plane and
starting with a random demagnetized state one has
$\rho(0)=\frac{1}{2}\rho_{//}+\frac{1}{2}\rho_{\bot}$. If the film
has uniaxial anisotropy ($\pm\emph{\textbf{M}}_{S}$ domains) one
then simply has $\rho(0)=\rho_{//}$. The so called anisotropic
magnetoresistance ratio (AMR) is given by \cite{AMR}:
\begin{equation}\label{Definição-de-AMR2}
AMR=\frac{\rho_{\parallel}-\rho_{\perp}} {\rho (0)}.
\end{equation}

\section{Experimental Results}
\label{Experimentalresults}
\subsection{MOKE magnetometry and imaging}

\begin{figure}
    \centering
    \includegraphics[width=0.80\textwidth]{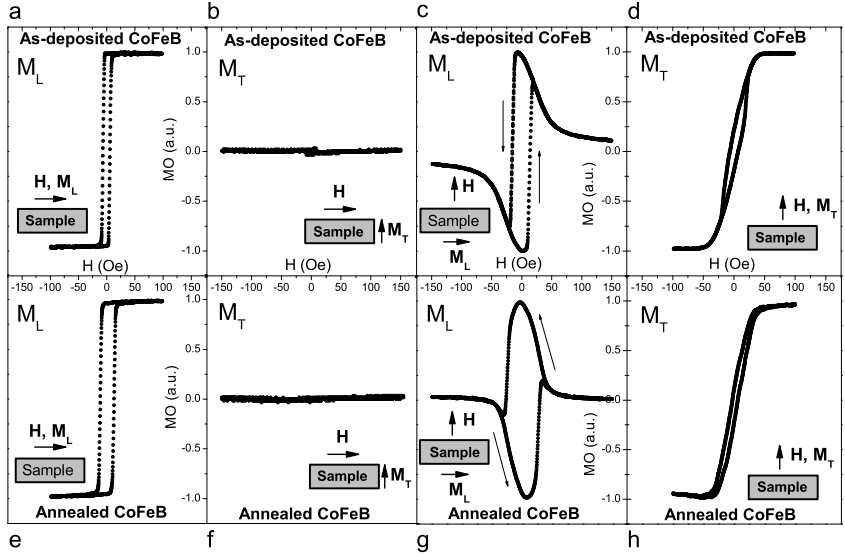}
    \caption{MOKE hysteretic cycles obtained for the annealed and
as-deposited CoFeB films. Each graph has a corresponding scheme
associated with the MOKE geometry. The (a)-(d) graphs were
obtained for the as-deposited film, while (e)-(h) correspond to
the annealed film.} \label{MOKEcycles}
\end{figure}

The magnetic loops obtained for the Co$_{73.8}$Fe$_{16.2}$B$_{10}$
thin films (as-deposited and annealed) are presented in Fig.
\ref{MOKEcycles}. The uniaxial anisotropy impressed during film
growth is readily apparent, having the easy (hard) axis oriented
parallel (perpendicular) to the light scattering plane defined by
the incident laser beam and the film normal. The easy direction
coincides with the long axis of the rectangular films.

For \emph{\textbf{H}} parallel to the easy axis we observe, in the
as-deposited CoFeB film (Fig. \ref{MOKEcycles}a), a rectangular
hysteretic cycle with a coercive field $H_c \simeq$ 477 Am$^{-1}$,
whereas $H_c\sim$ 955 Am$^{-1}$ for the annealed film (Fig.
\ref{MOKEcycles}e) i.e. the amorphous state reduces the coercive
field by a factor of $\sim$ 2. The saturation magnetic field
(\textit{$H_s$}; taken where irreversibility vanishes and a
constant magnetization plateau sets in) is also greatly reduced,
from $H_s\sim$ 1990 Am$^{-1}$ for the annealed film to $H_s\sim$
1273 Am$^{-1}$ in the amorphous CoFeB film. As expected, a zero
transverse magnetization component is measured in both films under
a longitudinal magnetic field (Figs. \ref{MOKEcycles}b and
\ref{MOKEcycles}f).

For \emph{\textbf{H}} perpendicular to the easy axis, typical
quasi-linear transverse-component $M_{T}(H)$ cycles are observed
as displayed in Figs. \ref{MOKEcycles}d and \ref{MOKEcycles}h.
However, small coercive fields are still observed, of 557
Am$^{-1}$ and 318 Am$^{-1}$ for the as-deposited and annealed
films respectively, leading to very narrow hysteretic cycles; the
saturation field is virtually the same in both cases, $H_{s} \sim
3980$ Am$^{-1}$ (here $H_{s}$ is taken where the magnetization
plateau sets in; for \textbf{\emph{H}} along the hard direction,
this occurs above the irreversibility point). On the other hand,
the longitudinal magnetization component in both films ($M_{L}$,
Figs. \ref{MOKEcycles}c and \ref{MOKEcycles}g) exhibits a small
field dependence on the approach to saturation, except for the
sudden magnetization reversal at $H_{c} \sim$ 1194 Am$^{-1}$ and
2228 Am$^{-1}$, for the as-deposited and annealed CoFeB films
respectively. On the other hand, $H_{s} \sim$ 7960 Am$^{-1}$ in
both cases.

\begin{figure}
    \centering
    \includegraphics[width=0.60\textwidth]{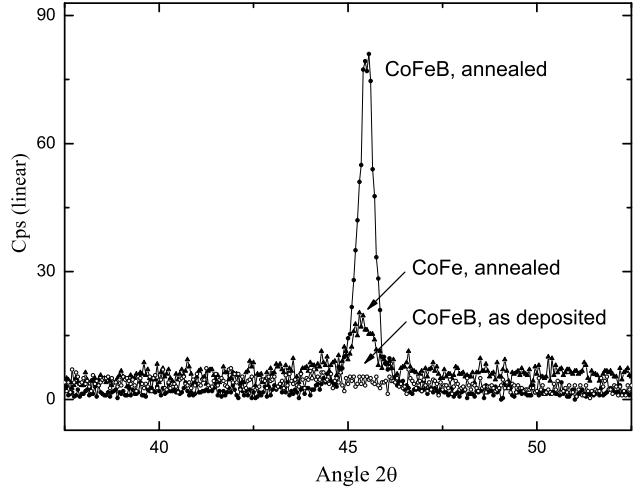}
   \caption{X-rays measurements of Co$_{80}$Fe$_{20}$ and Co$_{73.8}$Fe$_{16.2}$B$_{10}$ 20 nm thick films.
    \cite{CoFeB}, as deposited and annealed at $280^{\circ}$C. The as deposited CoFeB film is amorphous, whereas after $280^{\circ}$C anneal it becomes crystalline.}\label{raiox}
\end{figure}

Comparing these results with those previously obtained in annealed
Co$_{80}$Fe$_{20}$ films \cite{Teixeira}, we notice that the
introduction of boron in CoFe (favouring amorphous/nanocrystalline
structures; Fig. \ref{raiox}) reduces significatively the coercive
and saturation fields along the easy axis ($H_{c} \sim 2150$
Am$^{-1}$ and $H_{s} \sim 3980$ Am$^{-1}$ in Co$_{80}$Fe$_{20}$).
Also, the magnetization along the hard axis saturates at lower
fields in the CoFeB films (both as-deposited and annealed) than in
the annealed CoFe film ($H_{s} \sim$ 5970 Am$^{-1}$).

\begin{figure}
    \centering
    \includegraphics[width=1\textwidth]{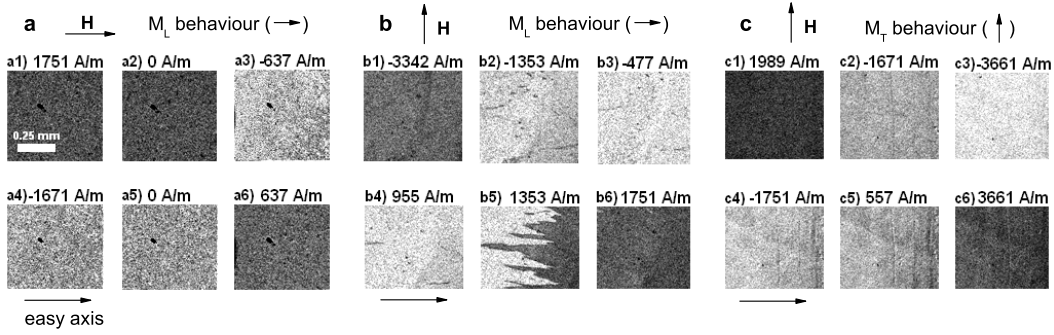}
    \caption{Magneto-optical Kerr images of as-deposited CoFeB thin film.
(a) The external field H was applied along the easy axis (parallel
to the light incident plane); images on the longitudinal ($M_{L}$)
magnetization component versus decreasing H (a1-a3) starting with
positive saturation (see also Fig. \ref{MOKEcycles}a); (b)
Magnetic field perpendicular to the easy axis (also parallel to
light incident plane); images on the longitudinal ($M_{L}$)
magnetization versus H (b1-b6), forward direction (Fig.
\ref{MOKEcycles}c); (c) External field perpendicular to the easy
axis (also parallel to light incident plane); images on the
$M_{T}(H)$ behaviour (c1-c3), backward direction (Fig.
\ref{MOKEcycles}d).}\label{MOKEimagesasdeposited}
\end{figure}

\begin{figure}
    \centering
    \includegraphics[width=1\textwidth]{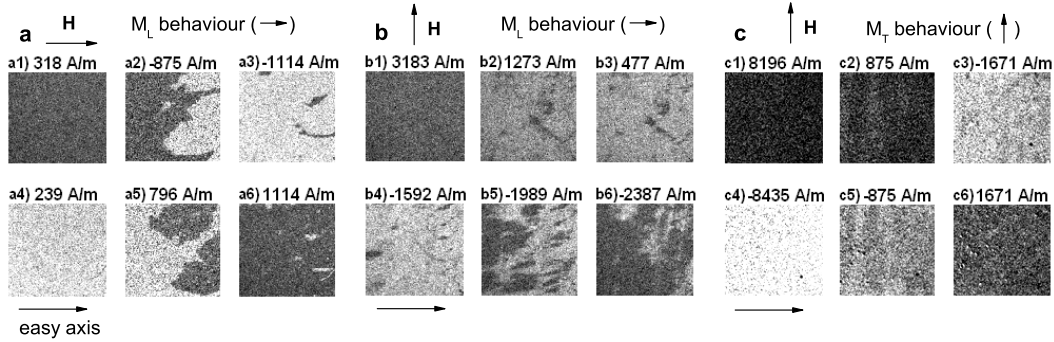}
    \caption{Magneto-optical images of annealed CoFeB thin film.
(a) The external magnetic field H was applied along the easy axis,
also parallel to the light incident plane; images on the
longitudinal ($M_{L}$) magnetization component versus H (a1-a3),
backward direction (Fig. \ref{MOKEcycles}e); (b) External magnetic
field perpendicular to the easy axis (also parallel to light
incident plane); images on the longitudinal ($M_{L}$)
magnetization versus H (b1-b6), backward direction (Fig.
\ref{MOKEcycles}g); (c) External field perpendicular to the easy
axis (also parallel to light incident plane); images on the
$M_{T}(H)$ behaviour (c1-c3), backward direction (Fig.
\ref{MOKEcycles}h).}\label{MOKEimagesannealed}
\end{figure}

The differences observed in the $M(H)$ curves for the two in-plane
MOKE geometries (longitudinal and transverse) are a consequence of
the two main reversal processes, rotation and domain wall
propagation \cite{Bozorth}. This was confirmed by visualization of
the magnetic domains, as summarized in Figs.
\ref{MOKEimagesasdeposited} and \ref{MOKEimagesannealed} for the
as-deposited and annealed CoFeB films respectively (see also
sections 5.1 and 5.2).

\subsection{AMR measurements}

\begin{figure}
    \centering
    \includegraphics[width=0.8\textwidth]{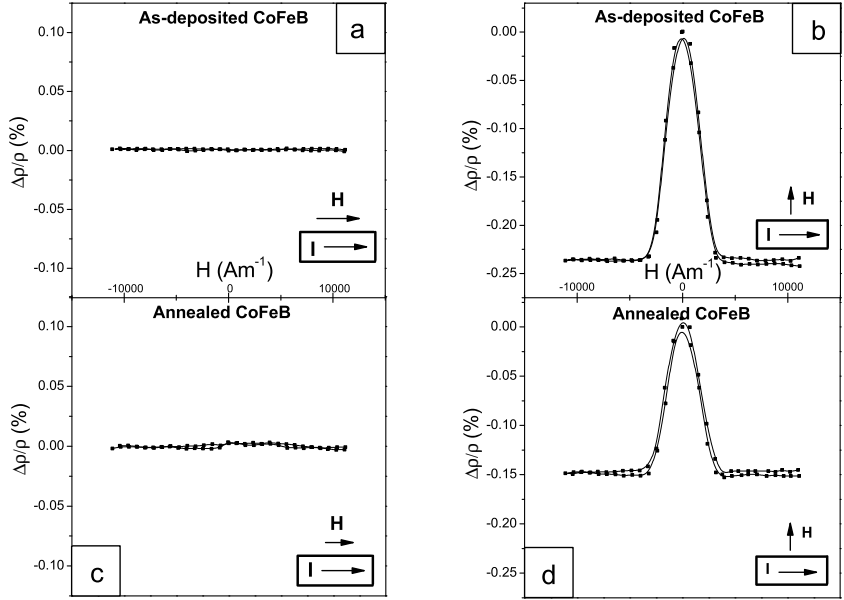}
    \caption{Magnetoresistance $\Delta\rho/\rho$ versus H in the as-deposited and annealed CoFeB thin films.
    (a) and (c) H parallel to the current (I); (b) and (d) H
    perpendicular to the current.}\label{AMR}
\end{figure}

The magnetoresistance $\Delta\rho/\rho$ vs $H$ curves with
\emph{\textbf{H}} parallel and perpendicular to the electrical
current \cite{AMR} are shown in Fig. \ref{AMR}. The
magnetoresistance is zero when the current and magnetic field are
parallel (Figs. \ref{AMR}a, \ref{AMR}c), and negative for the
as-deposited and annealed films when they are perpendicular
($\Delta\rho/\rho \sim -0.24\%$ and $-0.15\%$; Figs. \ref{AMR}b,
\ref{AMR}d). An AMR ratio of $\sim$ 0.24\% and $\sim$ 0.15\%
results from these later data, for the as-deposited and annealed
films respectively.

For the transverse configuration one can correlate the
$\Delta\rho/\rho(H)$ behaviour with that observed in M(H), while
under parallel magnetic fields no correlation is possible since MR
= 0 (see section \ref{AMRversusM(H)behavior}).

\section{Numerical analysis of $M(H)$ behaviour}

A simple treatment based on the Stoner-Wolfarth model
satisfactorily describes the qualitative behaviour of the
magnetization when the orientation of the magnetic field is
changed. According to this model, coherent magnetization rotation
is assumed. The relevant magnetic energy per unit volume ($E$) is
written as the sum of Zeeman ($E_{z}$) and anisotropy ($E_{a}$;
anisotropy constant $k_{1}$) energies:
\begin{equation}\label{Definicao-energia}
E=E_{Z}+E_{a}=-\mu_{0}\mathbf{M}\cdot\mathbf{H}+k_{1}\sin^{2}\theta
\end{equation}

\begin{figure}
    \centering
    \includegraphics[width=0.80\textwidth]{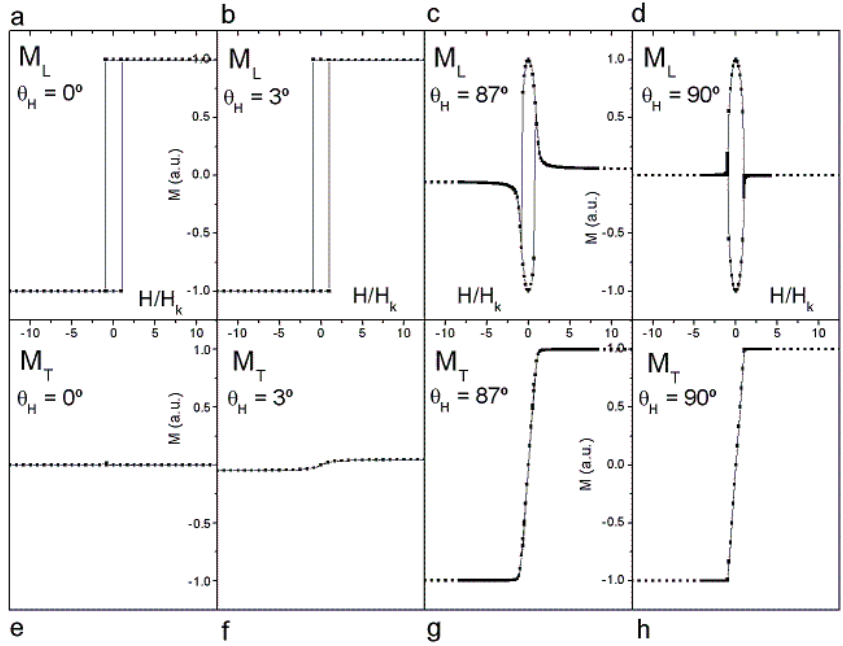}
    \caption{Numerical results of the $M_{L}$ and $M_{T}$ components for different orientations
    of the magnetic field relatively to the easy axis.
    (a) and (e) $\theta_{H} \sim 0^{\circ}$; (b) and (f) $\theta_{H} \sim 3^{\circ}$;
    (c) and (g) $\theta_{H} \sim 87^{\circ}$; (d) and (h) $\theta_{H} \sim 90^{\circ}$.}
    \label{Numericalresults}
\end{figure}

For an applied (in-plane) magnetic field making an angle
$\theta_{H}$ with the easy axis, and a magnetization making an
angle $\theta$ with the same axis, one can write:
\begin{equation}\label{Definicao-energia3}
E=-\mu_{0}M_{s}H\cos(\theta_{H}-\theta) + k_{1}\sin^{2}\theta
\end{equation}
We numerically obtain the $\theta$ angle which minimizes E (for
each value of H with a fixed $\theta_{H}$ orientation) and then
calculate the magnetization components $M_{L}=M_{s}\cos{\theta}$
and $M_{T}=M_{s}\sin{\theta}$. The results are illustrated in
Fig.\ref{Numericalresults}, for $\theta_{H} \sim 0^{\circ}$,
$3^{\circ}$, $87^{\circ}$ and $90^{\circ}$. The magnetic field is
normalized by the anisotropy field $H_{k} =
2k_{1}$/$\mu_{0}M_{S}$, where $M_{s}$ is the spontaneous
magnetization.

One sees that when \emph{\textbf{H}} is not exactly parallel or
perpendicular to the easy axis ($\theta_{H} \neq 0, 90^{\circ}$),
the magnetization processes become more complex. For small
$\theta_{H}$ values, i.e. small \emph{\textbf{H}} deviations from
the easy axis, the $M_{L}$ and $M_{T}$ hysteretic cycles are
hardly affected, because the small transverse magnetic field
component occurs along the hard direction for which the magnetic
susceptibility is rather low.

In contrast with this situation, for $\theta_{H} \sim 90^{\circ}$
a small field misalignment produces a field component along the
easy axis, which induces a significant $M_{L}$ component due to
the high magnetic susceptibility along such direction. A more
detailed analysis is given in next section.

\section{Analysis of $M_{T}(H)$ and $M_{L}(H)$ behaviour}
\label{AnalysisMTMLbehavior}

\subsection{As-deposited CoFeB film}

The rectangular easy-axis magnetization curve in Fig.
\ref{MOKEcycles}a, for the as-deposited CoFeB film, clearly
indicates an abrupt magnetization switching, which is directly
confirmed by the corresponding domain images in the backward (Fig.
\ref{MOKEimagesasdeposited}a1$\rightarrow$\ref{MOKEimagesasdeposited}a3)
and forward branches (Figs.
\ref{MOKEimagesasdeposited}a4$\rightarrow$\ref{MOKEimagesasdeposited}a6)
of the $M_{L}(H)$ loop. These sequences demonstrate a sudden
transition from one saturated magnetic state (single domain; dark
image) to the opposite state (single domain; bright image),
corresponding to a $180^{\circ}$ reversal of the magnetic domains
through domain wall motion.

When the magnetic field is applied along the hard axis (Fig.
\ref{MOKEcycles}d) the growth of the transverse magnetization is
quite gradual (almost linear-like), as also revealed by the
intensity of the MOKE images (Figs.
\ref{MOKEimagesasdeposited}c1$\rightarrow$\ref{MOKEimagesasdeposited}c6),
indicating the progressive rotation of the magnetization. On the
other hand, the $M_{L}(H)$ component (Fig. \ref{MOKEcycles}c)
indicates that the magnetization undergoes a $360^{\circ}$
rotation in the plane of the film, always in the same sense. This
process has two distinct regimes: one leading to a gradual
$M_{L}(H)$ dependence associated with domain rotations ($\sim
90^{\circ}$), and the other producing abrupt steps in the
magnetization, attributed to $\sim 180^{\circ}$ domain-wall
displacements (switching). As shown in the images in Figs.
\ref{MOKEimagesasdeposited}b1$\rightarrow$\ref{MOKEimagesasdeposited}b3
for the backward branch, initially (high fields) one observes a
progressive increase in the image brightness (proportional to
$M_{L}$), associated with rotation processes, while Figs.
\ref{MOKEimagesasdeposited}b4$\rightarrow$\ref{MOKEimagesasdeposited}b6
evidence abrupt changes in the image intensity (from bright to
dark) under small fields, due to the sudden magnetization
switching (180$^{\circ}$). Thus, with \textit{H} along the hard
axis we detect both magnetization reversal processes, with
dominance of magnetic domain rotations under high fields, and
sudden M-switching at low fields (domain wall motions).

We show numerically that when the magnetic field is perfectly
aligned with the hard axis (Figs. \ref{Numericalresults}d and
\ref{Numericalresults}h), the expected (Stoner-Wolfarth model)
magnetization process from one saturated state to another occurs
only by coherent moment rotations, with no sudden M-switching.
However, due to an experimentally unavoidable small
\emph{\textbf{H}}-misalignments with the hard axis, and incoherent
magnetization processes in real films, the two magnetization
processes (domain wall displacements and rotations) are
experimentally observed. The discontinuities ("jumps") given by
the Stoner-Wolfarth model, when there is a misalignment or when
the magnetic field is parallel to the easy-axis
(Figs.\ref{Numericalresults}a, \ref{Numericalresults}c,
\ref{Numericalresults}e and \ref{Numericalresults}g), are
physically due to domain wall motions.

\subsection{Annealed CoFeB film}

For the applied field along the easy axis, the MOKE results for
the annealed CoFeB film reveal again a rectangular magnetization
loop (Fig. \ref{MOKEcycles}e). Thus, the single domain
magnetization is switched from one direction to the opposite in an
extremely small field range, due to 180$^{\circ}$ domain-wall
formation and fast wall propagation, with no transverse
magnetization component ($M_T$; Fig. \ref{MOKEcycles}f).

\begin{figure}
    \centering
    \includegraphics[width=0.65\textwidth]{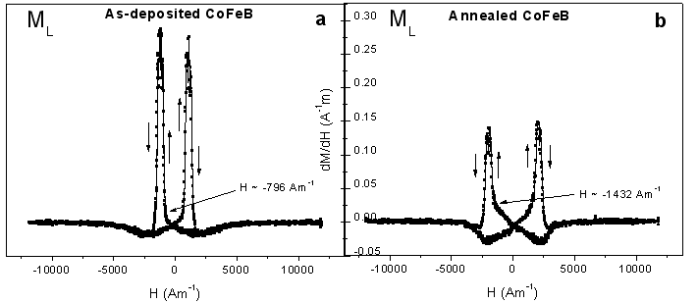}
    \caption{a) Field-derivatives of the $M_{L}$ component shown in Fig.\ref{MOKEcycles}c for the as-deposited CoFeB film.
    b) Derivatives of the $M_{T}$ component illustrated in Fig.\ref{MOKEcycles}g for the annealed CoFeB film.}
    \label{Derivatives}
\end{figure}

With the magnetic field along the hard axis (Figs.
\ref{MOKEcycles}g and \ref{MOKEcycles}h), we find again that the
magnetization process occurs through domain wall motion (switching
at low fields) and magnetic moment rotations. The domain wall
displacements are detected by the $M_L$ component (Fig.
\ref{MOKEcycles}g), through the rapid change in $M_L(H)$ within a
small field range. The much more gradual $M_T(H)$ dependence (Fig.
\ref{MOKEcycles}h), and also in some parts of $M_L(H)$ (Fig.
\ref{MOKEcycles}g), is attributed to domain rotations. Magnetic
domain visualization shows a complex pattern (Fig.
\ref{MOKEimagesannealed}b5), with nucleation and propagation of
magnetic domains through domain walls of reduced wall angle (less
than 180$^{\circ}$), since the variation of intensity between the
bright and dark regions is not so strong as previously observed
for the as-deposited CoFeB film with \emph{\textbf{H}} along the
hard axis (Fig. \ref{MOKEimagesasdeposited}b5). A careful analysis
of the $M_L(H)$ curve, using the highly sensitive and numerically
obtained $dM/dH$ derivatives, shows that such domain wall
propagation begins well after the $M_{L}$ maximum
(Fig.\ref{Derivatives}b; sharp peak in $dM_{L}/dH$ at $H \sim
-1430$ Am$^{-1}$ for the backward branch), contrarily to the
situation observed in the as-deposited CoFeB sample
(Fig.\ref{Derivatives}a; $dM_{L}/dH$ peak at $H \sim -795$
Am$^{-1}$ for the backward branch), indicating that some of the
magnetic moments leave the easy axis.

Again, a qualitatively good correlation between the numerical
results and the experimental ones is observed. As above, due to an
small \emph{\textbf{H}}-misalignments with the hard axis, both
reversal processes are present. The discontinuities given by the
Stoner-Wolfarth model (Figs.\ref{Numericalresults}a,
\ref{Numericalresults}c, \ref{Numericalresults}e and
\ref{Numericalresults}g) are real and physically related to domain
wall displacements.

\section{AMR versus M(H) behaviour}
\label{AMRversusM(H)behavior}

The magnetoresistance measurements show that $\Delta\rho/\rho$ is
very sensitive to the type of processes underlying the
magnetization reversal.

Under transverse magnetic fields, when magnetization rotation
dominates, we have similar saturation magnetic fields in the
$M(H)$ and $(\Delta\rho/\rho)(H)$ curves. In this geometry, for
which domain rotations dominate, we indeed expect important
variations in the resistivity with magnetic field through the Smit
mechanism (Eq. (\ref{Definição-de-R})). At low (transverse) fields
the longitudinal magnetization component displays a maximum and
the magnetoresistance is very small, since the magnetic moments
are essentially aligned along the easy axis ($\theta \approx 0$ or
$\pi$; M$_T$/M$_L$ $\ll$ 1) and, according to Eq.
(\ref{Definição-de-R}), $\rho(H)\sim\rho_{//}\sim\rho_0$ which
leads to $\Delta\rho/\rho \sim 0$.

Under longitudinal magnetic fields, the magnetization loops and
magnetic domain images of Figs. \ref{MOKEcycles}a,
\ref{MOKEcycles}e, \ref{MOKEimagesasdeposited}a and
\ref{MOKEimagesannealed}a indicate $180^{\circ}$ magnetization
switching at low fields, and consequently no variation in the
electrical resistivity (Fig. \ref{AMR}; in agreement with Eq.
(\ref{Definição-de-R}), since $\cos^{2}\theta = \cos^{2}(\theta +
\pi)$).

In conclusion, amorphous and nanocrystalline CoFeB films show low
coercive fields and abrupt switching, as well as absence of short
range spin disorder effects after switching (which were observed
in crystalline CoFe thin films \cite{Teixeira}).

Work supported in part by FEDER/POCTI/0155, IST-2001-37334 NEXT
MRAM and FEDER/POCTI/CTM/59318/2004 project. J. Ventura, R.
Ferreira and S. Cardoso are thankful for the FCT grants
SFRH/BD/7028/2001, SFRH/BD/6501/2001 and SFRH/BPD/7177/2001
respectively.


\begin{thebibliography}{11}

\bibitem{Modera}J. S. Moodera, L. R. Kinder, T. M. Wong and R. Meservey, \emph{Phys.
Rev. Lett}. {\bf 74}, 3273 (1995).
\bibitem{readheads}D. Song, J. Nowak, R. Larson, P. Kolbo, R. Chellew,
\emph{IEEE Trans. Magn.} \textbf{36} 2545 (2000).
\bibitem{MRAMs}S. Tehrani, B. Engel, J. M. Slaughter, E. Chen, M. DeHerrera,
M. Durlam, P. Naji, R. Whig, J. Janesky and J. Calder, \emph{IEEE
Trans. Magn.} \textbf{36} 2752 (2000).
\bibitem{CoFeB70}D. Wang, C. Nordman, J. M. Daughton, Z. Qian and J. Fink,
 \emph{IEEE Trans. Magn.} \textbf{40} 2269 (2004).
\bibitem{thermalstability}T. Dimopoulos, G. Gieres, J. Wecker, N. Wiese and M. D. Sacher,
 \emph{J. Appl. Phys.} \textbf{96} 6382 (2004).
\bibitem{CoFeB}S. Cardoso, R. Ferreira, P. P. Freitas, M. MacKenzie, J. Chapman, J. O. Ventura,
J. B. Sousa and U. Kreissig, \emph{IEEE Trans. Magn.} \textbf{40}
2272 (2004).
\bibitem{Teixeira} J. M. Teixeira, R.F.A. Silva, J. Ventura, A.
Pereira, J. P. Ara\'ujo, M. Amado, F. Carpinteiro, J. B. Sousa, S.
Cardoso, R. Ferreira and P. Freitas, accepted to \emph{Materials
Science Forum}.
\bibitem{AMR}T. R. McGuire and R. I. Potter, \emph{IEEE Trans. Magn.} {\bf
11}, 1018 (1975).
\bibitem{Bozorth}Richard M. Bozorth, \emph{Ferromagnetism}, IEEE Press
(1951).
\bibitem{AlexHubert}Alex Hubert and Rudolf Sch\"{a}fer, \emph{Magnetic
Domains; The Analysis of Magnetic Microstructures}, Springer
(1998).





\end{thebibliography}
\end{document}